# Integrating Emotion Distribution Networks and Textual Message Analysis for X User Emotional State Classification

Pardis Moradbeiki
*Department of Electrical and Computer Engineering*
Isfahan University of Technology
Isfahan, Iran
p.moradbeiki@ec.iut.ac.ir

Mohammad Ali Zare Chahooki
*Department of Electrical and Computer Engineering*
Yazd University
Yazd, Iran
chahooki@yazd.ac.ir

**Abstract**: As the popularity and reach of social networks continue to surge, a vast reservoir of opinions and sentiments across various subjects inundates these platforms. Among these, X social network (formerly Twitter) stands as a juggernaut, boasting approximately 420 million active users. Extracting users' emotional and mental states from their expressed opinions on social media has become a common pursuit. While past methodologies predominantly focused on the textual content of messages to analyze user sentiment, the interactive nature of these platforms suggests a deeper complexity. This study employs hybrid methodologies, integrating textual analysis, profile examination, follower analysis, and emotion dissemination patterns. Initially, user interactions are leveraged to refine emotion classification within messages, encompassing exchanges where users respond to each other. Introducing the concept of a communication tree, a model is extracted to map these interactions. Subsequently, users' bios and interests from this tree are juxtaposed with message text to enrich analysis. Finally, influential figures are identified among users' followers in the communication tree, categorized into different topics to gauge interests. The study highlights that traditional sentiment analysis methodologies, focusing solely on textual content, are inadequate in discerning sentiment towards significant events, notably the presidential election. Comparative analysis with conventional methods reveals a substantial improvement in accuracy with the incorporation of emotion distribution patterns and user profiles. The proposed approach yields a 12% increase in accuracy with emotion distribution patterns and a 15% increase when considering user profiles, underscoring its efficacy in capturing nuanced sentiment dynamics.

Code and Dataset: [Github](Github)

## 1. Introduction:

In the contemporary digital landscape, social networks have emerged as vital conduits for communication, information dissemination, and social interaction. Among these platforms, Twitter stands out as a dynamic microblogging platform where millions of users engage daily, expressing a diverse array of opinions, emotions, and sentiments on various topics. The pervasive nature of social media has led to an exponential increase in the volume of user-generated content, presenting both opportunities and challenges for researchers seeking to understand human behavior and sentiment in online environments.

Understanding the emotional states of individuals on social media platforms like Twitter is not only of academic interest but also holds practical implications across diverse domains such as marketing, politics, and public health. The ability to accurately classify and analyze emotions expressed in social media posts can provide valuable insights into public opinion, sentiment trends, and emerging societal issues. However, this task is inherently complex due to the dynamic nature of language, the nuances of context, and the sheer volume of data generated in real-time.

Traditionally, sentiment analysis methods have focused primarily on analyzing the textual content of messages to infer the underlying sentiment or emotional tone. While these approaches have yielded valuable insights, they often fail to capture the intricacies of human emotion expressed in social media interactions. Social networks are inherently interactive platforms where users engage in conversations, share content, and influence each other's opinions and emotions through their interactions. Thus, a deeper understanding of user sentiment requires consideration of both textual content and the broader social context in which it is situated.

In this paper, we propose a novel approach to classify the emotional states of Twitter users by integrating the analysis of textual messages with insights derived from users' profiles, interactions, and the propagation of emotions within the network. We leverage the rich contextual information available on Twitter, including user bios, interests, follower relationships, and patterns of interaction, to enhance the accuracy and granularity of emotion classification. By combining these diverse sources of data, we aim to capture the multifaceted nature of human emotion expressed in social media conversations.

The remainder of this paper is structured as follows: in Section 2, we provide an overview of related work in the field of sentiment analysis and emotion detection on social media. In Section 3, we describe the methodology employed in our approach, outlining the various data sources and techniques utilized for emotion classification. Section 4 presents the experimental setup and results of our study, followed by a discussion of the findings in Section 5. Finally, we conclude the paper in Section 6, summarizing our contributions. Through this work, we seek to advance our understanding of emotional expression on social media platforms and contribute to the development of more robust and accurate sentiment analysis techniques.

## 2. related work

The realm of sentiment analysis and emotion detection in social media has garnered substantial attention from researchers across various disciplines due to its profound implications for understanding human behavior, opinion dynamics, and societal trends. In this section, we review existing literature that has explored methodologies, techniques, and insights relevant to our study on classifying emotional states of Twitter users using a hybrid approach.

Early research in sentiment analysis predominantly focused on lexicon-based methods, wherein sentiment scores were assigned to individual words or phrases based on predefined dictionaries of sentiment-bearing terms. These approaches laid the groundwork for sentiment analysis but often struggled to capture the nuances of emotion expression in social media text due to the contextual and dynamic nature of language.

To address these limitations, researchers began to explore machine learning and natural language processing (NLP) techniques for sentiment analysis on social media platforms. Supervised learning algorithms, such as Support Vector Machines (SVM), Naive Bayes, and deep learning models like Recurrent Neural Networks (RNNs) and Convolutional Neural Networks (CNNs), were employed to classify text into sentiment categories (positive, negative, neutral).

In recent years, there has been a shift towards more context-aware and domain-specific sentiment analysis approaches tailored to the unique characteristics of social media data. Some studies have integrated domain-specific lexicons and sentiment resources to improve sentiment classification accuracy, leveraging domain knowledge and linguistic cues specific to social media discourse.

Additionally, researchers have explored the role of user-level features, such as social network structure, user interactions, and user profiles, in shaping emotional expression and sentiment patterns on social media platforms. Several studies have demonstrated the utility of incorporating user-level features to enhance emotion detection and sentiment analysis, highlighting the importance of considering the broader social context in which messages are situated.

Moreover, researchers have investigated the propagation of emotions within social networks and the influence of social network structure on emotional dynamics. Emotion diffusion models and network-based approaches have been developed to model the spread of emotions and identify influential users or communities that drive emotional contagion within the network.

Furthermore, recent studies have explored the integration of multimodal data sources (e.g., text, images, videos) for emotion analysis, recognizing the richness and diversity of content shared on social media platforms. By incorporating multiple modalities, researchers aim to capture a more holistic understanding of emotional expression in online environments.

There are two primary approaches to analyzing aggregated documents. The first involves bottom-up methods, which commence with individual models and subsequently extrapolate them to draw conclusions about the entire population [2]. The composition of the sample under scrutiny holds significant importance as it often solely represents the principal categories, crucial for the analysis. To ensure a representative sample of the population, researchers have employed various corrective techniques, including the stratified sampling method [3], simple random sampling [4], and multi-stage random sampling [5] [1].

On the other hand, the second approach comprises top-down methods, which contrast with the preceding methods by utilizing cumulative population data for inference [6]. Previous studies on users' posts and metadata have indicated that demographic details can be discerned using various machine learning algorithms, yielding accuracies ranging from 60% to 90% [7], or from profile information like username, page name, biography, and profile picture, with a Macro F1 score of 0.9 for gender and a Macro F1 score of 0.5 for age being attainable [8].

Sentiment analysis within user posts remains a prominent focus of research. Previous studies [10] [9] have illustrated the significance of emotions and emotional states in influencing individuals' interactions with technology.

Recent investigations have identified features indicative of user behavior, categorizing them into various groups such as post frequency, posting time patterns, and follower counts. Peng et al. segmented these features into user profile, behavior, and textual elements, employing a multi-core support vector machine for classification [11]. Social network attributes, including tweet count and interaction frequency, are leveraged, alongside personal information sourced from user profiles.

User-level characteristics, derived from aggregate Twitter activities like tweet and retweet counts, are often considered alongside linguistic styles for feature extraction. Resnik et al. introduced a model to uncover nuanced structures within tweets [12], while Shen et al. examined user behavior across Twitter and Weibo [14], utilizing linguistic features for sentiment analysis, employing the TextMind system for Chinese language analysis.

Differing from post-level features extracted from individual tweets, user-level features are derived from multiple tweets over time [14]. Chen et al. proposed a deep emotion representation model integrating CNN and LSTM, leveraging CNN layers for text feature extraction and LSTM for content comprehension. This model, combined with a one-versus-one training approach, outperformed SVM, CNN, LSTM, and CNN-LSTM models, achieving 78.42% accuracy in multi-category sentiment analysis [15].

User profile attributes, posting habits, and interaction metrics like follower counts are examined. Similarly, Weng et al. combined user-level and post-level features, formulating their analysis as a multi-class learning task, capitalizing on user-level tags to enhance post-level tag identification [16].

Yang et al. introduced SLCABG, a sentiment analysis model merging sentiment lexicons with Convolutional Neural Networks (CNN) and Bidirectional Attention-Based Recurrent Units (BiGRU). This model combines sentiment dictionaries with deep learning, addressing limitations of prior models in product review sentiment analysis.

Salver et al. proposed a hybrid deep learning model incorporating various word embedding methods (Word2Vec, FastText, letter-based embedding) and deep learning architectures (LSTM, GRU, BiLSTM, CNN). This model aggregates features from diverse deep learning techniques for sentiment categorization, outperforming previous studies with an 82% accuracy rate in emotion classification [17].

### 3. proposed model:

The proposed model in this paper addresses the challenge of sarcasm in social network messages by recognizing the insufficiency of analyzing text alone. Furthermore, considering the hierarchical structure of information flow in social networks, the model suggests examining message text within a tree structure to ascertain emotional polarity.

Outlined in Figures 1-3, the proposed method entails utilizing comments, messages, and follower profile data to construct the tree. Nodes in the tree are labeled using two key methods:

1. Methods grounded in emotional lexicons to gauge the sentiment level of messages.

2. Recommender system techniques to assess the similarity between user profiles.

Additionally, sentiment information from Twitter users is integrated into the labeling process. The final labeling step employs a Multi-Layer Perceptron (MLP) to compute the ultimate label.

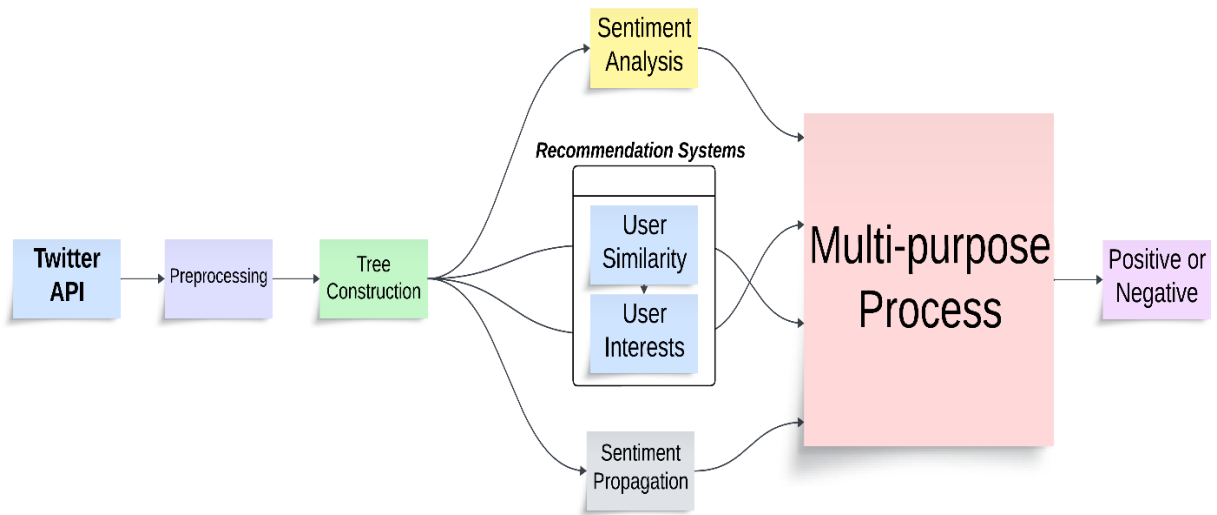

Fig. 1-3: proposed model

## 3.1. Preprocessing

Preprocessing serves as a pivotal phase in readying data for analysis, offering the potential to enhance model accuracy and efficiency through varied techniques. Textual data undergoes cleansing, whereby special characters, numbers, and punctuation marks are removed to streamline recognition. Standardizing text to lowercase fosters consistency and diminishes data dimensions. Notably, expunging specific URLs, mentions, and hashtags associated with social media platforms is imperative. Text tokenization aids in segmenting text into manageable units such as words or phrases, facilitating analysis. Depending on analytical requisites, methods like word tokenization or nth tokenization, which records word sequences, are employed. The elimination of commonplace words, termed stop words (e.g., "a," "this," "is"), holds paramount significance and can be executed via predefined or customized lists based on data and domain expertise.

Rooting and word stemming methods aim to condense words to their base or stem form, enhancing computational efficiency. The choice between these techniques hinges on analysis-specific requirements. Managing negative words and emoticons assumes significance, as their presence profoundly influences sentence sentiment and meaning. Emoticons or emojis can be treated as distinct entities or omitted altogether.

Addressing misspelled words entails employing spell-checking methodologies and libraries for spell correction, while expanding abbreviations ensures accurate text representation. Furthermore, numerical features can be scaled or normalized to a standardized range (e.g., 0 to 1) to mitigate the dominance of certain features during model training. In instances of skewed positive and negative sample distributions, techniques like oversampling, downsampling, or class weighting can rectify class imbalance.

In the proposed method, diverse data sources converge to construct a comprehensive tree structure, encompassing:

1. 60 messages sourced from Twitter and retweets.

2. User profiles, inclusive of status updates, follower lists, account creation timestamps, and user locations.

3. Top 100 members recommended by Twitter.

Following preprocessing, the subsequent step involves constructing a dependency propagation tree.

## 3.2. Dependency tree

As previously noted, a dependency tree is crafted for each user to evaluate the sentiment polarity of a message. As the distance or edges between two nodes increase, the mutual influence between these nodes diminishes, underscoring the significance of tree depth as a measure of effectiveness. Each node within the tree encapsulates a collection of user messages pertinent to the subject of inquiry, with both parent messages and profile information serving as tags for message categorization.

The dependency tree is represented as a triple $T(V,E,I)$, where $V$ denotes a node, $E$ signifies an edge, and $I$ represents a polarity label. Each node corresponds to a tweet or retweet, while each edge denotes a retweet. In the proposed approach, the maximum depth of impact within the tree is set at 4.

Emotion dissemination patterns play a crucial role in sentiment classification, particularly when analyzing social media data. These patterns describe how emotions spread and evolve within online interactions. They consider factors like:

1. *Contagion*: Positive or negative emotions can be contagious, influencing the sentiment of others who encounter them.
2. *Echo Chambers*: Users tend to connect with others who share similar views, reinforcing existing emotions within these groups.
3. *Cascading Effects*: A single event or post can trigger a cascade of emotional responses, amplifying the initial sentiment.
4. *Influencer Impact*: Prominent users or accounts can significantly influence the emotions expressed by their followers.

How They Improve Sentiment Classification:

1. *Contextual Understanding*: By incorporating dissemination patterns, sentiment analysis can go beyond the surface meaning of words. It can understand how emotions are influenced by the conversational context and social network dynamics.
2. *Identifying Trends*: Dissemination patterns can help identify emerging trends in emotions. For example, a sudden surge in negative sentiment might indicate a brewing public relations crisis.
3. *Distinguishing Genuine Sentiment*: These patterns can help differentiate genuine sentiment from manipulation. For example, a coordinated effort to spread negativity might be identifiable through analyzing dissemination patterns.

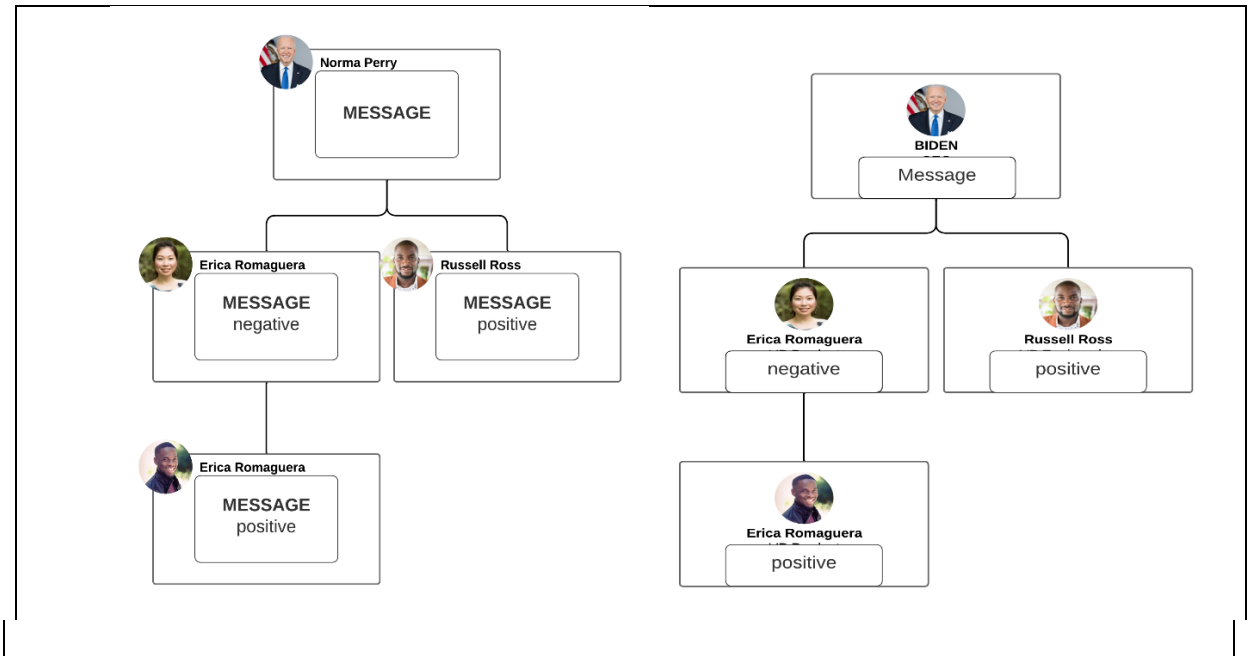

Fig.2 Dependency tree structure

Two methods are employed for tagging the dependency tree:

1. *Word-based sentiment analysis*:

In this approach, two databases, WordNet and SentiWordNet [19], are utilized. WordNet comprises words and their corresponding roots, while SentiWordNet assigns a polarity score to each word. For each root word and its synonyms, the database is queried, and a score is assigned to this collection based on the SentiWordNet database.

2. *User interests*:

To discern the interests of each user [20], the topics of influential users on Twitter are initially extracted. These influential individuals are ranked based on user interest since users can freely follow them.

 2.1. *Topic modeling*:

This modeling leverages Wikipedia data to extract information about famous individuals, including their areas of interest. Using Latent Dirichlet Allocation (LDA), word clusters are generated based on topics, with each word assigned a weight between 0 and 1. The output indicates the interest level of each influential Twitter user regarding a particular topic.

2.2. *Building a ranking matrix*:

The similarity of two users' sentiments towards a subject depends on their overall similarity. Common followers between two individuals are suggested as a similarity parameter. The similarity calculation involves averaging the number of shared users between two nodes within the dependency tree.

2.3. *CatBoost*:

The CatBoost classification algorithm is employed to categorize users' interests into predetermined groups. Unlike conventional methods like one-hot encoding, CatBoost avoids producing sparse matrices and overfitting. This algorithm utilizes a binary tree as the foundational predictor in gradient approximation. It optimizes parameters through a greedy approach.

3. *Cascade tree for narrowly deep branching diffusion*:

In Political Analysis, Analyzing how political messages evoke and spread emotions can provide insights into public opinion.

This research investigates the impact of a novel pattern extracted from user interaction trees on sentiment analysis. We leverage a pre-trained model, trained on a massive dataset of messages, to initially assign sentiment labels to individual messages.

Our proposed inference pattern works by traversing the dependency tree from the bottom (leaf nodes) upwards towards the root. We start with the sentiment assigned by the pre-trained model for a leaf node. To determine the sentiment of the parent message in the tree, we consider the tree's height:

- *Height 1*: If the tree has only one level (leaf and root), the sentiment of the leaf node is directly transferred to the parent.
- *Height 2*: If the tree has two levels (leaf, middle node, and root), the leaf node's sentiment is compared to the middle node's sentiment.
    - If both the leaf and middle node are positive, the parent message receives a positive sentiment.
    - If the leaf is positive but the middle node is negative, or vice versa, the parent receives a negative sentiment.

This process continues for trees with more levels, comparing the sentiment of the leaf node to the sentiment of the intermediate node closest to the root. The results indicate that most user interaction trees have a height of 4 or 5 [21]. We therefore considered patterns up to height 5 for extracting the sentiment data.

The analysis demonstrates that this diffusion pattern, which takes into account message interactions, leads to a more accurate understanding of how emotions spread and evolve online compared to traditional sentiment analysis methods.

### 3.3. MLP

The last stage involves labeling the messages based on the two aforementioned methods, each selecting a distinct label.

### 4. Evaluation

For this study, we require three distinct datasets. The initial dataset comprises 60 published messages, the second dataset encompasses 100 profiles of renowned individuals featured on Twitter, and the third dataset comprises profiles of individuals on Twitter. Tables 1-4, 2-4, and 3-4 exhibit the sample datasets correspondingly. Subsequently, a detailed explanation of the dataset employed is provided. Furthermore, a dataset comprising 143,000 messages concerning the American elections of 2020, sourced from www.kaggle.com, was utilized. Each message within this dataset is assigned one of three labels: positive, negative, or neutral.

Tables 4.1. messages for dependency tree

| Tweet_ID | Text | Retweet_ID | Label |
|---|---|---|---|
| *optus* | *When we took office, we didnot waste a second getting the pandemic under control and our economy back on track. Just five months later, we administered over 300 million shots and unemployment is at its lowest level since the pandemic started. America is coming back.* | *null* | *positive* |
| *mubazieric* | *The economy created more than 1.3 million new jobs in 100 days. That more new jobs in the first 100 days than any president on record. Promises delivered, That's a good start.* | *optus* | *positive* |
| *Juditho65763855* | *I'm so grateful the Biden Administration has made getting the pandemic under control such a priority. It has given us our lives and economy back.* | *optus* | *positive* |
| *pearlyB57* | *Build Back Better!* | *optus* | *negative* |
| *waitingOnBiden* | *We never got the $2,000 checks, nationwide mask mandate, Medicare expansion, or public option you promised, and you're set to fail your July 4th vaccination goal.* | *optus* | *negative* |

Tables 4.2. dataset of fames X member

| Name | Twitter_ID | Followers | Activity |
|---|---|---|---|
| *Barack Obama* | *BarackObama* | *11,08,90,048* | *Politician* |
| *Katy perry* | *Katyperry* | *10,83,15,414* | *Musician* |
| *Justin bieber* | *justinbieber* | *10,74,10,873* | *Musician* |
| Rihanna | rihanna | 9,49,90,708 | Musician and Businesswomen |
| Taylor Swift | Taylorswift13 | 8,55,20,236 | Musician |

Tables 4.3. profile X member

| ID | Name | bio | Location |
|---|---|---|---|
| POTUS | President Biden | 46th President of the United States, husband to @FLOTUS , proud dad & pop. Tweets may be archived: http://whitehouse.gov/privacy | United states |
| mubazieric | ERIC JEM | Team Biden, Entrepreneur, Cyber Security professional,Christian, Democrat , Public speaker, #HumanRights, Vaccinated #ClimateAction #Resist #WearAMask | New York |
| Juditho65763855 | Judith Olson | Resist. Love my family. So proud of President Biden and his Administration. Get vaccinated, wear a mask, save your life. No D.M.'s unless I know you. | null |
| pearlyB57 | PearlyBVaccinated | Librarian, knitter, gardener, Mom. Blue state dweller; lucky me! #GunSenseVoter #VoteBlue #BidenHarris I'm here to learn the truth. | The Other Washington |
| waitingOnBiden | Holding Biden Accountable | Bad things are still bad when Dems do them. Biden controls Congress & can cancel debt by EO. Tracking how "nothing would fundamentally change." By @westonpagano | Washington, DC |

## 5. Evaluation Metric

To assess the proposed method, we've scrutinized it against various criteria, including precision, accuracy, recall, and F score.

For the evaluation, we start by defining four key factors: FN (False Negatives), FP (False Positives), TN (True Negatives), and TP (True Positives). TP represents the count of correctly predicted positive samples, while TN indicates the count of correctly predicted negative samples. FP signifies the count of falsely predicted positive samples, and FN represents the count of falsely predicted negative samples. Utilizing these factors, we compute precision, accuracy, recall, and F score.

Accuracy, one of the evaluation metrics, quantifies the ratio of correct predictions to the total number of test data points. It essentially elucidates how many samples are accurately classified among the entire test dataset. While accuracy is commonly used for gauging prediction model performance due to its simplicity, it may not always be the most suitable criterion for assessing classification algorithms. Hence, to ensure a comprehensive evaluation of the proposed model, we incorporate additional criteria alongside accuracy. The accuracy evaluation criterion is derived using equations 1-4.

$$Accuracy = \frac{TP + TN}{TP + TN + FP + FN}$$

(Equation 5-1)

The Precision criterion represents the proportion of positive predictions made by the algorithm. Essentially, it quantifies the ratio between accurate positive predictions and the total positive predictions. This concept is depicted in Figure 5-1. The evaluation of accuracy is calculated using Equation 5-2.

$$Precision = \frac{TP}{TP + FP}$$

(Equation 5-2)

Another practical measure is recall, which assesses the ability to predict the number of positive samples correctly. The calculation of the recall evaluation criterion is derived from Equation 5-3.

$$Recall = \frac{TP}{TP + FN}$$

(Equation 5-3)

Another valuable metric is the F score, which represents the harmonic mean of accuracy and recall. It typically indicates the model's suitability with the data. The F score evaluation criterion is computed using equation 5-4.

$$F - measure = 2.\left(\frac{Precision.recall}{precision + recall}\right)$$

(Equation 5-4)

The crucial and impactful factors affecting the proposed model have been delineated and documented in Table 5-1. These values are derived either from standard references cited in articles [44] or from validation data.

Tables 5.1. The crucial and impactful factors

| | |
|---|---|
| Considered LDA Topic | 19 |
| Learning rate of catboost | 1,2 |
| Tree depth | 4,…,10 |
| Iteration of catboost | 2000 |

Based on the findings presented in Table 5-2, the process of trial and error has yielded the most optimal outcome.

Tables 5.2. Evault

| | *Tree depth=10 Iteration=2000* | *Tree depth=10 Iteration=2000* | *Tree depth=16 Iteration=2000* | *Tree depth=16 Iteration=2000* |
|---|---|---|---|---|
| *Accuracy* | 0.6290 | 0.6129 | 0.6451 | 0.6451 |
| *Precision* | 0.6086 | 0.5869 | 0.6417 | 0.7150 |
| *Recall* | 0.5789 | 0.5718 | 0.5859 | 0.5664 |
| *F-measure* | 0.5698 | 0.5679 | 0.5694 | 0.5163 |

Previous studies solely focused on the textual content of messages, disregarding user communication. In this section, we contrast the impact of the additional factors with the proposed model, based on evaluation criteria. The evaluation outcomes are displayed in 5-3.

Tables 5.3. The impact of the supplementary factors incorporated in the proposed approach

| | *Lexicon_based* | *Proposed_model* |
|---|---|---|
| *Accuracy* | 0.5230 | 0.6451 |
| *Precision* | 0.5625 | 0.7150 |
| *Recall* | 0.5500 | 0.5664 |
| *F-measure* | 0.5088 | 0.5163 |

The comparative chart illustrating the proposed method alongside each specified factor is depicted in Figure 5-1. As illustrated, each of the three factors enhances the evaluation criteria, potentially boosting accuracy by 12% and precision by 14%.

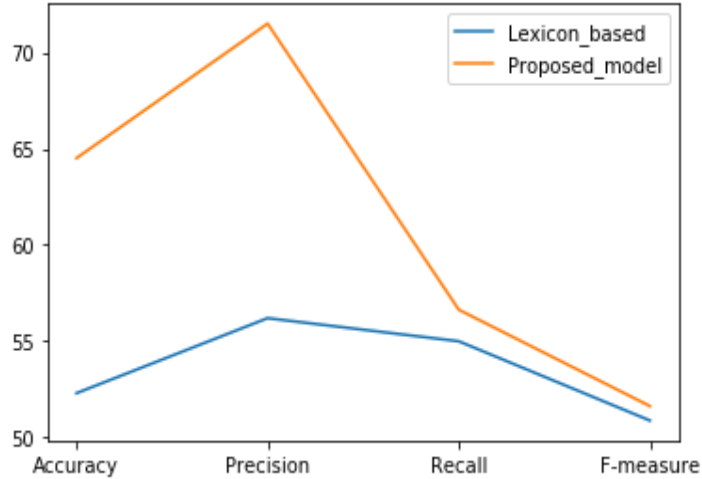

Fig. 5.1. Evaluation of the supplemental factors in the proposed method

For evaluation, messages exchanged during the 2020 American elections were employed. Initially, our model was constructed utilizing machine learning approaches such as Newby's methods, neural networks, and support vector machine method, followed by comparison of the resultant evaluation criteria. Table 5-4 presents the comparative results of these methods vis-à-vis the proposed method.

Furthermore, the comparative chart of these methods with the proposed approach is depicted in Figure 5-2. Additionally, in Figure 5-3, we juxtapose the evaluation criteria incorporating the supplementary factors of the proposed method against conventional methods.

Table 5.4. Contrast between the proposed method and conventional approaches

|  | *Accuracy* | *Precision* | *Recall* | *F-measure* |
|---|---|---|---|---|
| *Tf-idf + Naïve bayse* | *0.714* | *0.441* | *0.384* | *0.374* |
| *Tf-idf + Neural network* | *0.634* | *0.396* | *0.449* | *0.410* |
| *Tf-idf + SVM* | *0.698* | *0.434* | *0.494* | *0.454* |
| *Tf-idf + SVM + User_interested + Sentiment diffusion* | *0.619* | *0.612* | *0.603* | *0.602* |

| Tf-idf + SVM + User_similarity + User_interested + Sentiment diffusion | 0.639 | 0.6202 | 0.604 | 0.615 |
|---|---|---|---|---|
| Proposed_model | 0.6451 | 0.715 | 0.566 | 0.516 |

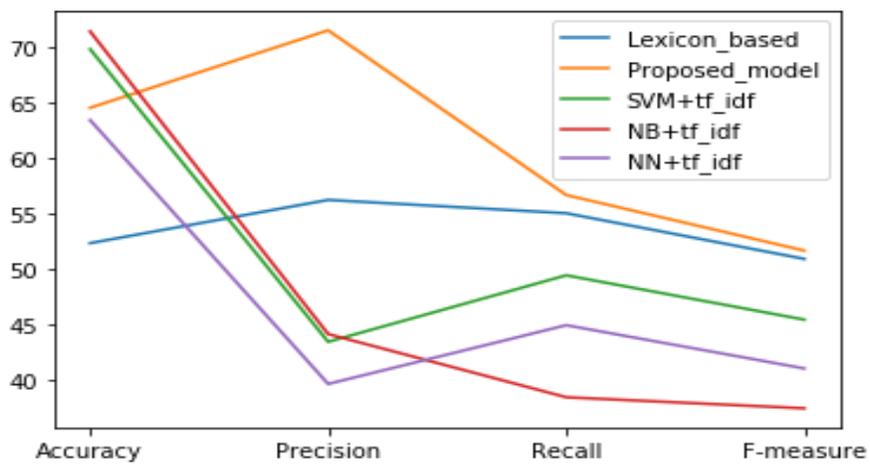

Fig. 5.2. Comparison chart illustrating the proposed method alongside conventional approaches

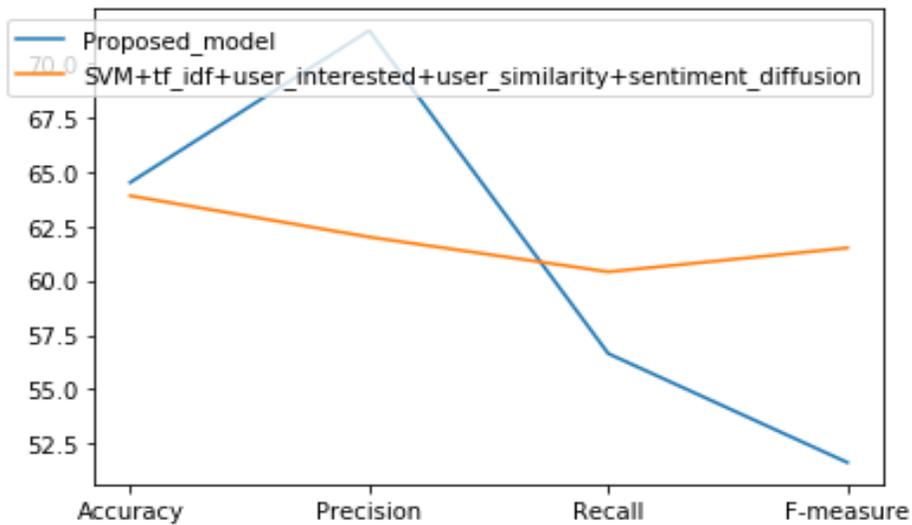

Fig. 5.3. Comparison chart showcasing the suggested factors by conventional methods alongside the proposed method

The results demonstrate that the proposed method achieves higher precision and accuracy.

## 6. Conclusion

Following the preparation of the requisite dataset, a tree structure was constructed based on tweets and retweets. This research comprises three main segments. Firstly, textual content from messages is employed to gauge emotional sentiment. Secondly, utilizing information gleaned from users' profiles, the similarity of each user to a designated root user is computed, potentially a prominent figure. This segment utilizes content-based approaches akin to recommender systems. Thirdly, each user's profile information is leveraged to discern their preferred topics, derived from the famous individuals they follow and the interests of fellow users. This stage combines content-based approaches with collaborative filtering within the realm of recommender systems. Lastly, a cascade tree is utilized to discern patterns, facilitating the comparison of each individual with the root node. This segment delves into information dissemination patterns within a deep tree structure.